\documentclass[12pt,a4paper,aps,preprint,superscriptaddress,nofootinbib]{revtex4-1}
\usepackage[utf8]{inputenc}
\usepackage{graphicx}
\usepackage{amssymb}
\usepackage{textcomp}
\usepackage{amsmath}
\usepackage{tabularx}
\usepackage{bm}
\usepackage{times}
\usepackage{color}
\usepackage{slashed}
\usepackage{multirow}
\usepackage{verbatim}
\usepackage{cancel}
\usepackage{subfigure}
\usepackage{ulem}

\usepackage[colorlinks=true, pdfstartview=FitV, linkcolor=blue, citecolor=blue, urlcolor=blue]{hyperref}
\allowdisplaybreaks[4]

\def\lsim{\mathrel{\raise.3ex\hbox{$<$\kern-.75em\lower1ex\hbox{$\sim$}}}}
\def\gsim{\mathrel{\raise.3ex\hbox{$>$\kern-.75em\lower1ex\hbox{$\sim$}}}}

\definecolor{orange}{rgb}{1,0.5,0}

\begin{document}

\title{Searching for high-frequency axion in quantum electromagnetodynamics through interface haloscopes}

\author{Tong Li}
\email{litong@nankai.edu.cn}
\affiliation{
School of Physics, Nankai University, Tianjin 300071, China
}
\author{Chang-Jie Dai}
\email{daichangjie@mail.nankai.edu.cn}
\affiliation{
School of Physics, Nankai University, Tianjin 300071, China
}
\author{Rui-Jia Zhang}
\email{zhangruijia@mail.nankai.edu.cn}
\affiliation{
School of Physics, Nankai University, Tianjin 300071, China
}

\begin{abstract}
The so-called
Witten effect implies a close relationship between axion and magnetic monopole. A sound quantization in the presence of magnetic monopoles, called quantum electromagnetodynamics (QEMD), was utilized to construct a more generic axion-photon Lagrangian in the low-energy axion effective field theory. This generic axion-photon Lagrangian introduces the interactions between axion and two four-potentials, and leads to new axion-modified Maxwell equations. The interface haloscopes place an interface between two electromagnetic media with different properties and are desirable to search for high-mass axions $m_a\gtrsim \mathcal{O}(10)~\mu{\rm eV}$. In this work, for the generic axion-photon couplings built under QEMD, we perform comprehensive calculations of the axion-induced propagating waves and energy flux densities in different interface setups. We also obtain the sensitivity to new axion-photon couplings for high-mass axions.
\end{abstract}

%\arxivnumber{arXiv:2304.12525}

%\begin{document}

\maketitle
%\setcounter{page}{2}

%\newpage

%%%%%%%%%%%%%%%%%%%%%%%%%%%%%%%%
\section{Introduction}
\label{sec:Intro}
%%%%%%%%%%%%%%%%%%%%%%%%%%%%%%%%

The strong CP problem in quantum chromodynamics (QCD) arises from the severe constraint on the Chern-Simons $\theta$ term as source of CP violation from neutron EDM measurement~\cite{Baluni:1978rf,Crewther:1979pi,Kim:1979if,Shifman:1979if,Dine:1981rt,Zhitnitsky:1980tq,Baker:2006ts,Pendlebury:2015lrz}. The most famous solution is the Peccei-Quinn (PQ) mechanism with a QCD anomalous $U(1)_{\rm PQ}$ global symmetry~\cite{Peccei:1977hh,Peccei:1977ur,Weinberg:1977ma,Wilczek:1977pj}. The spontaneous breaking of $U(1)_{\rm PQ}$ introduces a pseudo-Goldstone boson $a$ called axion. It induces a coupling $-g_{a\gamma\gamma} a F^{\mu\nu}\tilde{F}_{\mu\nu}/4=g_{a\gamma\gamma} a \vec{E}\cdot \vec{B}$ between axion and electromagnetic fields in QED. The other motivation for axion is that it can make up the total dark matter (DM) density during phase transition in the early universe~\cite{Preskill:1982cy,Abbott:1982af,Dine:1982ah}. The reasonable mass range of axion for a cold DM takes as $m_a\approx \mathcal{O}(10)~\mu{\rm eV}$.
The success of such DM axion paradigm pivots on the search for the conversion of axion into electromagnetic field in cavity haloscope experiments~\cite{Sikivie:1983ip} such as ADMX~\cite{ADMX:2021nhd}. They look for the axion-induced radiation as a solution of axion-modified Maxwell equations, suppose the resonance is tuned to the right axion mass. Theoretically, the axion mass is not limited on the above ADMX sensitive range. It is also desirable to search for DM axion over the broader possible mass range~\cite{Sikivie:2020zpn}. For instance, $m_a\lesssim \mathcal{O}(1)~\mu{\rm eV}$ can be searched though electronic LC circuit~\cite{Sikivie:2013laa} such as ABRACADABRA~\cite{Kahn:2016aff,Salemi:2021gck} and ADMX SLIC~\cite{Crisosto:2019fcj} or high-mass range of $m_a\gtrsim \mathcal{O}(10)~\mu{\rm eV}$ through dish antenna~\cite{Horns:2012jf} or dielectric haloscopes~\cite{Millar:2016cjp} such as MADMAX~\cite{Caldwell:2016dcw}.

In 1979, E.~Witten pointed out that a CP violating term in the non-Abelian $SO(3)$ theory provides an additional electric charge for the ’t Hooft-Polyakov monopoles in this theory~\cite{Witten:1979ey}. This is the so-called
Witten effect which implies the existence of relationship between axion and magnetic monopole.
Later on, W.~Fischler et al. derived this axion-dyon dynamics under the classical electromagnetism~\cite{Fischler:1983sc}. In order to describe the axion-dyon dynamics in quantum field theory (QFT), Ref.~\cite{Sokolov:2022fvs} recently constructed a more generic axion-photon Lagrangian in the low-energy axion effective field theory (EFT).
They utilized a reliable quantization in the presence of magnetic monopoles developed by J.~S.~Schwinger and D.~Zwanziger in 1960's, called quantum electromagnetodynamics (QEMD)~\cite{Schwinger:1966nj,Zwanziger:1968rs,Zwanziger:1970hk}. This generic low-energy axion-photon Lagrangian accounts for the Witten effect term as well and introduces three more interesting interactions between axion and two four-potentials~\cite{Sokolov:2022fvs,Sokolov:2023pos}. This axion EFT Lagrangian also leads to new axion-modified Maxwell equations~\cite{Sokolov:2022fvs,Li:2022oel,Sokolov:2023pos}.

Very recently, there exist quite a few theoretical and phenomenological works on the generic axion-photon interactions under QEMD~\cite{Li:2022oel,Tobar:2022rko,Sokolov:2023pos,Li:2023aow,Tobar:2023rga}. Ref.~\cite{Li:2022oel} properly solved the new axion-modified Maxwell equations and proposed new LC strategies to measure the new couplings for sub-$\mu$eV axion. A more recent paper Ref.~\cite{Tobar:2023rga} also studied the LC regime. Refs.~\cite{Tobar:2022rko,Li:2023aow} applied the Poynting theorem or quantum calculation to determine how
to obtain the sensitivity to new axion-photon couplings for $\mu{\rm eV}$ axions in resonant haloscopes. However, the formulas and detection strageties of new axion-photon couplings are still lacking for high-mass range of axions $m_a\gtrsim \mathcal{O}(10)~\mu{\rm eV}$ based on interface haloscopes. The traditional interface haloscopes place a disc (or multiple discs) between which there are two dielectric media with different permittivity in a background magnetic field. The axion-induced electric field on either side jumps at the interface surface. Then, a propagating wave is produced to satisfy the continuous boundary conditions and thus emits in both perpendicular directions to the interface surface. The more reliable dielectric haloscopes measure the energy flux density of the propagating waves by setting a perfect mirror as well as multiple interfaces~\cite{Millar:2016cjp,Caldwell:2016dcw}.
In this work, for the generic axion-photon couplings built under QEMD, we provide comprehensive expressions for the axion-induced electromagnetic fields and the propagating waves in different interface setups. We also apply the Poynting's theorem to calculate the energy flux densities and obtain the sensitivity to new axion-photon couplings for high-mass axions.

This paper is organized as follows. In Sec.~\ref{sec:QEMDandMaxwell}, we introduce the generic axion-photon interactions in QEMD and derive the new axion modified Maxwell equations. Their macroscopic form and the linearization of fields and medium response are also performed. In Sec.~\ref{sec:Solution}, we obtain the axion-induced radiation at an interface between two regions of different media. Then, we utilize the Poynting's theorem to obtain the energy flux density. Possible new interface haloscopes are discussed in Sec.~\ref{sec:Exp}. We also show the numerical results of high-frequency axion search potentials in terms of such interface haloscopes. Our conclusions are drawn in Sec.~\ref{sec:Con}.

%%%%%%%%%%%%%%%%%%%%%%%%%%%%%%%%%%%%%%%%%
\section{The modified Maxwell equations of axion in QEMD}
\label{sec:QEMDandMaxwell}
%%%%%%%%%%%%%%%%%%%%%%%%%%%%%%%%%%%%%%%%%

%%%%%%%%%%%%%%%%%%%%%%%%%%%%%%%%%%%%%%%%%
\subsection{The generic axion-photon interactions in QEMD}
\label{sec:QEMD}
%%%%%%%%%%%%%%%%%%%%%%%%%%%%%%%%%%%%%%%%%

The QEMD framework introduces two four-potentials $A^\mu$ and $B^\mu$ to describe photon, instead of one four-potential in the standard electromagnetism. The corresponding $U(1)$ gauge group of QEMD is replaced by $U(1)_{\rm E}\times U(1)_{\rm M}$ whose conserved charges are electric and magnetic charges. Based on the QEMD theory, a generic low-energy axion-photon EFT can be built~\cite{Sokolov:2022fvs}.
The Lagrangian for the generic interactions between axion $a$ and photon in QEMD is~\footnote{We follow Ref.~\cite{Tobar:2023rga} to change the notation of couplings to $g_{aEE}$ ($=g_{a\gamma\gamma}$), $g_{aEM}$, and $g_{aMM}$. They are equivalent to $g_{aAA}$, $g_{aAB}$, and $g_{aBB}$ in Ref.~\cite{Sokolov:2022fvs}, respectively.}~\cite{Sokolov:2022fvs}
\begin{eqnarray}
\mathcal{L}&=& {1\over 2n^2} \{[n\cdot (\partial \wedge B)]\cdot [n\cdot (\partial \wedge \tilde{A})] -[n\cdot (\partial \wedge A)]\cdot [n\cdot (\partial \wedge \tilde{B})] - [n\cdot (\partial\wedge A)]^2 \nonumber \\
&&- [n\cdot (\partial\wedge B)]^2\} -{1\over 4} g_{aEE}~a~{\rm tr}[(\partial \wedge A)(\partial \wedge \tilde{A})] - {1\over 4} g_{aMM}~a~{\rm tr}[(\partial \wedge B)(\partial \wedge \tilde{B})]\nonumber \\
&& - {1\over 2} g_{aEM}~a~{\rm tr}[(\partial \wedge A)(\partial \wedge \tilde{B})] -j_e \cdot A - j_m\cdot B +\mathcal{L}_G \;,
\end{eqnarray}
where $(\partial \wedge X)^{\mu \nu}\equiv \partial^\mu X^\nu - \partial^\nu X^\mu$ for four-potential $X^\mu=A^\mu$ or $B^\mu$, $(\partial \wedge \tilde{X})^{\mu \nu}\equiv \epsilon^{\mu\nu\rho\sigma} (\partial \wedge X)_{\rho\sigma}/2$ as the Hodge dual tensor with $\epsilon_{0123}=+1$, $n^\mu = (0,\vec{n})$ is an arbitrary fixed spatial vector, and $\mathcal{L}_G$ is a gauge-fixing term. Note that we ignore the term for Witten effect here.
The electromagnetic field strength tensors $F^{\mu\nu}$ and $\tilde{F}^{\mu\nu}$ are then introduced in the way that
\begin{eqnarray}
F=\partial \wedge A - (n\cdot \partial)^{-1} (n\wedge \tilde{j}_m)\;,~~
\tilde{F}=\partial \wedge B + (n\cdot \partial)^{-1} (n\wedge \tilde{j}_e)\;,
\label{eq:FFd}
\end{eqnarray}
where $j_e$ and $j_m$ are electric and magnetic currents, respectively. Thus, the two four-potentials have opposite parities.
The first two dimension-five operators ($g_{aEE}$ and $g_{aMM}$ terms) are CP-conserving axion interactions. Their couplings $g_{aEE}$ and $g_{aMM}$ are given by the $U(1)_{\rm PQ}U(1)_{\rm E}^2$ and $U(1)_{\rm PQ}U(1)_{\rm M}^2$ anomalies, respectively. Note that the coupling $g_{aEE}$ is equivalent to the standard coupling $g_{a\gamma\gamma}$. As $A^\mu$ and $B^\mu$ have opposite parities, the third operator ($g_{aEM}$ term) is a CP-violating one. Its coupling $g_{aEM}$ is determined by the $U(1)_{\rm PQ}U(1)_{\rm E}U(1)_{\rm M}$ anomaly.

The above coupling coefficients can be calculated as
\begin{eqnarray}
g_{aEE}={Ee^2\over 4\pi^2 v_{\rm PQ}}\;,~~g_{aMM}={Mg_0^2\over 4\pi^2 v_{\rm PQ}}\;,~~g_{aEM}={Deg_0\over 4\pi^2 v_{\rm PQ}}\;,
\label{eq:couplings}
\end{eqnarray}
where $v_{\rm PQ}$ is the $U(1)_{\rm PQ}$ symmetry breaking scale, $e$ is the unit of electric charge, and $g_0$ is the minimal magnetic charge with $g_0=2\pi/e$ under the Dirac-Schwinger-Zwanziger (DSZ) quantization condition. $E$ and $M$ are the electric and magnetic anomaly coefficients, respectively. $D$ is the coefficient from mixed electric-magnetic CP-violating anomaly. They can be computed by following Fujikawa's path integral method~\cite{Fujikawa:1979ay} and integrating out heavy PQ-charged fermions with electric and magnetic charges. As the DSZ quantization condition tells $g_0\gg e$, according to Eq.~(\ref{eq:couplings}), we have the hierarchy of the axion-photon couplings as $g_{aMM}\gg g_{aEM}\gg g_{aEE}$.

%%%%%%%%%%%%%%%%%%%%%%%%%%%%%%%%%%%%%%%%%
\subsection{The axion modified Maxwell equations and their macroscopic form}
\label{sec:Maxwell}
%%%%%%%%%%%%%%%%%%%%%%%%%%%%%%%%%%%%%%%%%

According to the above Lagrangian for generic axion-photon interactions, we can derive the classical equations of motion for the photon field. By applying the Euler-Lagrange equation of motion for the two potentials, one obtains
\begin{eqnarray}
&&{1\over n^2}(n\cdot\partial n\cdot\partial A^\mu - n\cdot\partial \partial^\mu n\cdot A-n\cdot\partial n^\mu \partial\cdot A-n\cdot\partial \epsilon^\mu_{\nu\kappa\lambda}n^\nu \partial^\kappa B^\lambda) \nonumber \\
&&-g_{aEE}\partial_\nu a (\partial\wedge \tilde{A})^{\nu\mu}-g_{aEM}\partial_\nu a (\partial\wedge \tilde{B})^{\nu\mu}= j_e^\mu\;,\\
&&{1\over n^2}(n\cdot\partial n\cdot\partial B^\mu - n\cdot\partial \partial^\mu n\cdot B-n\cdot\partial n^\mu \partial\cdot B+n\cdot\partial \epsilon^\mu_{\nu\kappa\lambda}n^\nu \partial^\kappa A^\lambda) \nonumber \\
&&-g_{aMM}\partial_\nu a (\partial\wedge \tilde{B})^{\nu\mu}-g_{aEM}\partial_\nu a (\partial\wedge \tilde{A})^{\nu\mu}= j_m^\mu\;.
\end{eqnarray}
In terms of the field strength tensors $F^{\mu\nu}$ and $\tilde{F}^{\mu\nu}$, the following axion modified Maxwell equations are obtained~\cite{Sokolov:2022fvs}
\begin{eqnarray}
&&\partial_\mu F^{\mu\nu} -g_{aEE} \partial_\mu a \tilde{F}^{\mu\nu} + g_{aEM} \partial_\mu a F^{\mu\nu} = j_e^\nu\;,  \\
&&\partial_\mu \tilde{F}^{\mu\nu} +g_{aMM} \partial_\mu a F^{\mu\nu} - g_{aEM} \partial_\mu a \tilde{F}^{\mu\nu} = j_m^\nu\;,
\end{eqnarray}
where the term responsible for Witten effect is omitted.
The new Maxwell equations in terms of electric and magnetic fields are then given by
\begin{eqnarray}
&&\vec{\nabla}\times \vec{B}-{\partial \vec{E}\over \partial t}=\vec{j}_e+g_{aEE}(\vec{E} \times \vec{\nabla} a - {\partial a\over \partial t} \vec{B})
+ g_{aEM} (\vec{B} \times \vec{\nabla} a + {\partial a\over \partial t} \vec{E})\;,
\label{eq:Ampere}
\\
&&\vec{\nabla}\times \vec{E}+{\partial \vec{B}\over \partial t}=\vec{j}_m-g_{aMM}(\vec{B} \times \vec{\nabla} a + {\partial a\over \partial t} \vec{E})
- g_{aEM} (\vec{E} \times \vec{\nabla} a - {\partial a\over \partial t} \vec{B})\;,\\
&&\vec{\nabla}\cdot \vec{B} = \rho_m -g_{aMM} \vec{E}\cdot \vec{\nabla} a + g_{aEM} \vec{B}\cdot \vec{\nabla} a \;,\\
&&\vec{\nabla}\cdot \vec{E} = \rho_e + g_{aEE} \vec{B}\cdot \vec{\nabla} a - g_{aEM} \vec{E}\cdot \vec{\nabla} a \;,
\end{eqnarray}
where the magnetic charge $\rho_m$ and current $\vec{j}_m$ will be ignored below as there is no observed magnetic monopole.

Next, we derive the macroscopic form of the above Maxwell equations in order to deal with the propagating wave in media. Let's first recall the classical electromagnetism. In media, both the electric charge and the current are composed of a free part and a bound part
\begin{eqnarray}
\rho_{e} = \rho_{e,f} + \rho_{e,b}\;,~~~\vec{j}_e = \vec{j}_{e,f} + \vec{j}_{e,b}\;.
\end{eqnarray}
where the bound parts are given by
\begin{eqnarray}
\rho_{e,b}=-\vec{\nabla}\cdot\vec{P}_e\;,~~~\vec{j}_{e,b}=\vec{\nabla}\times \vec{M}_e + {\partial \vec{P}_e\over \partial t}\;.
\end{eqnarray}
Here $\vec{P}_e$ and $\vec{M}_e$ denote the macroscopic polarization and magnetization, respectively. Moreover, the free parts satisfy the continuity equation
\begin{eqnarray}
{\partial \rho_{e,f}\over \partial t} + \vec{\nabla}\cdot\vec{j}_{e,f} =0\;.
\end{eqnarray}
The macroscopic electric displacement field $\vec{D}$ and the macroscopic magnetic field $\vec{H}$ are defined as
\begin{eqnarray}
\vec{D}=\vec{E}+\vec{P}_e\;,~~~\vec{H}=\vec{B}-\vec{M}_e\;.
\end{eqnarray}
After plugging them into QEMD Ampere's law Eq.~(\ref{eq:Ampere}), we obtain
\begin{eqnarray}
\vec{\nabla}\times \vec{H}-{\partial \vec{D}\over \partial t}=\vec{j}_{e,f}+g_{aEE}(\vec{E} \times \vec{\nabla} a - {\partial a\over \partial t} \vec{B})
+ g_{aEM} (\vec{B} \times \vec{\nabla} a + {\partial a\over \partial t} \vec{E})\;.
\end{eqnarray}
The axion-photon interaction terms are not affected by the medium response. Thus, the electromagnetic fields on the right-hand side of the above equation are $\vec{E}$ and $\vec{B}$ but not $\vec{D}$ and $\vec{H}$. Similarly, the QEMD Gauss's law can be rewritten as
\begin{eqnarray}
\vec{\nabla}\cdot \vec{D} = \rho_{e,f} + g_{aEE} \vec{B}\cdot \vec{\nabla} a - g_{aEM} \vec{E}\cdot \vec{\nabla} a \;.
\end{eqnarray}

There is no conventional source in Faraday's law and the divergence of $\vec{B}$. Their equations thus remain unaffected. Finally, we obtain the macroscopic form of axion Maxwell equations in QEMD
\begin{eqnarray}
&&\vec{\nabla}\times \vec{H}-{\partial \vec{D}\over \partial t}=\vec{j}_{e,f}+g_{aEE}(\vec{E} \times \vec{\nabla} a - {\partial a\over \partial t} \vec{B})
+ g_{aEM} (\vec{B} \times \vec{\nabla} a + {\partial a\over \partial t} \vec{E})\;,
\label{eq:mac1}
\\
&&\vec{\nabla}\times \vec{E}+{\partial \vec{B}\over \partial t}=-g_{aMM}(\vec{B} \times \vec{\nabla} a + {\partial a\over \partial t} \vec{E})
- g_{aEM} (\vec{E} \times \vec{\nabla} a - {\partial a\over \partial t} \vec{B})\;,
\label{eq:mac2}
\\
&&\vec{\nabla}\cdot \vec{B} = -g_{aMM} \vec{E}\cdot \vec{\nabla} a + g_{aEM} \vec{B}\cdot \vec{\nabla} a \;,
\label{eq:mac3}
\\
&&\vec{\nabla}\cdot \vec{D} = \rho_{e,f} + g_{aEE} \vec{B}\cdot \vec{\nabla} a - g_{aEM} \vec{E}\cdot \vec{\nabla} a \;.
\label{eq:mac4}
\end{eqnarray}
These are the macroscopic wave equations that we shall solve below.

%%%%%%%%%%%%%%%%%%%%%%%
\subsection{Decomposition and linearization}
%%%%%%%%%%%%%%%%%%%%%%%

Next, we decompose and linearize the fields in the above macroscopic axion Maxwell equations. Because the axion couplings are expected to be small, as a good perturbative approximation, we can expand the fields in Eqs.~(\ref{eq:mac1},\ref{eq:mac2},\ref{eq:mac3},\ref{eq:mac4}) (denoted by $\vec{X}=\vec{H},\vec{D},\vec{B},\vec{E}$) in terms of order of axion couplings. The decomposition is thus $\vec{X}=\vec{X}_0+\vec{X}_{\gamma a}$ where $\vec{X}_0$ is the large static electromagnetic fields and $\vec{X}_{\gamma a}\equiv\vec{X}_\gamma + \vec{X}_a$ corresponds to the homogeneous axion-induced fields $\vec{X}_a$ and the propagating waves $\vec{X}_\gamma$. The electric charge density $\rho_{e,f}$ and current density $\vec{j}_{e,f}$ can also be decomposed as the part causing the large background fields and the axion source term, i.e., $\rho_{e,f}=\rho_{e,f}^0+\rho'_{e,f}$ and $\vec{j}_{e,f}=\vec{j}_{e,f}^0+\vec{j}'_{e,f}$.
Thus, the static background fields satisfy the ordinary Maxwell equations
\begin{align}
\vec{\nabla}\times \vec{B}_{0}-{\partial \vec{E}_{0}\over \partial t}&=\vec{j}_{e,f}^0\;,
\label{eq:j0}\\
\vec{\nabla}\times \vec{E}_{0}+{\partial \vec{B}_{0}\over \partial t}&=0\;,\\
\vec{\nabla}\cdot \vec{B}_{0} &=0\;, \\
\vec{\nabla}\cdot \vec{E}_{0} &=\rho_{e,f}^0\;.
\label{eq:rho0}
\end{align}
Next, we assume that the static fields $\vec{B}_{0}$ and $\vec{E}_{0}$ are ideally provided and perfectly homogeneous. Thus, both Eqs.~(\ref{eq:j0}) and (\ref{eq:rho0}) turn to be approximately equal to zero, i.e., $\vec{\nabla}\times \vec{B}_{0}-{\partial \vec{E}_{0}\over \partial t}\approx 0$ and $\vec{\nabla}\cdot \vec{E}_{0}\approx 0$.
After keeping only the leading terms on the right-hand side, in terms of the static external electromagnetic fields $\vec{B}_0$ and $\vec{E}_0$, we have
\begin{eqnarray}
&&\vec{\nabla}\times \vec{H}-{\partial \vec{D}\over \partial t}=\vec{j}'_{e,f}+g_{aEE}(\vec{E}_0 \times \vec{\nabla} a - {\partial a\over \partial t} \vec{B}_0)
+ g_{aEM} (\vec{B}_0 \times \vec{\nabla} a + {\partial a\over \partial t} \vec{E}_0)\;,
\\
&&\vec{\nabla}\times \vec{E}+{\partial \vec{B}\over \partial t}=-g_{aMM}(\vec{B}_0 \times \vec{\nabla} a + {\partial a\over \partial t} \vec{E}_0)
- g_{aEM} (\vec{E}_0 \times \vec{\nabla} a - {\partial a\over \partial t} \vec{B}_0)\;,
\\
&&\vec{\nabla}\cdot \vec{B} = -g_{aMM} \vec{E}_0\cdot \vec{\nabla} a + g_{aEM} \vec{B}_0\cdot \vec{\nabla} a \;,
\\
&&\vec{\nabla}\cdot \vec{D} = \rho'_{e,f} + g_{aEE} \vec{B}_0\cdot \vec{\nabla} a - g_{aEM} \vec{E}_0\cdot \vec{\nabla} a \;.
\end{eqnarray}
Applying a time derivative on the last two equations gives
\begin{eqnarray}
&&\vec{\nabla}\cdot \dot{\vec{B}} = -g_{aMM} \vec{E}_0\cdot \vec{\nabla} \dot{a} + g_{aEM} \vec{B}_0\cdot \vec{\nabla} \dot{a} \;,\\
&&\vec{\nabla}\cdot (\dot{\vec{D}}+\vec{j}'_{e,f}) = g_{aEE} \vec{B}_0\cdot \vec{\nabla} \dot{a} - g_{aEM} \vec{E}_0\cdot \vec{\nabla} \dot{a} \;.
\end{eqnarray}
These equations are now linear for all times-space dependent quantities.
We then perform a Fourier expansion for the quantities in form of plane waves given by $e^{-i(\omega t - \vec{k}\cdot \vec{x})}$
\begin{eqnarray}
&&\vec{k}\times \hat{\vec{H}}+\omega \hat{\vec{D}} + i\hat{\vec{j}}'_{e,f}=g_{aEE}(\vec{E}_0\times \vec{k} \hat{a}+\omega \hat{a} \vec{B}_0)+g_{aEM}(\vec{B}_0\times \vec{k} \hat{a}-\omega \hat{a} \vec{E}_0)\;,\\
&&\vec{k}\times \hat{\vec{E}}-\omega \hat{\vec{B}} =-g_{aMM}(\vec{B}_0\times \vec{k} \hat{a}-\omega \hat{a} \vec{E}_0)-g_{aEM}(\vec{E}_0\times \vec{k} \hat{a}+\omega \hat{a} \vec{B}_0)\;,\\
&&\vec{k}\cdot \hat{\vec{B}} = -g_{aMM}\vec{k}\cdot \vec{E}_0 \hat{a} + g_{aEM}\vec{k}\cdot \vec{B}_0 \hat{a}\;,\\
&&\vec{k}\cdot (\omega\hat{\vec{D}} +i\hat{\vec{j}}'_{e,f}) = g_{aEE}\omega\vec{k}\cdot \vec{B}_0 \hat{a} - g_{aEM}\omega\vec{k}\cdot \vec{E}_0 \hat{a}\;,
\end{eqnarray}
where $\hat{a}$, $\hat{\vec{H}}$, $\hat{\vec{D}}$, $\hat{\vec{E}}$, $\hat{\vec{B}}$ and $\hat{\vec{j}}'_{e,f}$ are all amplitudes as a function of $\omega$ and $\vec{k}$.
%

%%%%%%%%%%%%%%%%%%%%%%%%%%%%%%%%%%%%%%%%%
\section{QEMD axion-induced radiation at an interface}
\label{sec:Solution}
%%%%%%%%%%%%%%%%%%%%%%%%%%%%%%%%%%%%%%%%%

\begin{figure}[h!]
\begin{center}
%\minigraph{7cm}{-0.05in}{(a)}{case1.png}
%\hspace{1cm}
%\minigraph{7cm}{-0.05in}{(b)}{case2.png}\\
%\vspace{1cm}
%\minigraph{7cm}{-0.05in}{(c)}{case3.png}
%\vspace{1cm}
%\minigraph{7cm}{-0.05in}{(d)}{case4.png}
\includegraphics[scale=0.6]{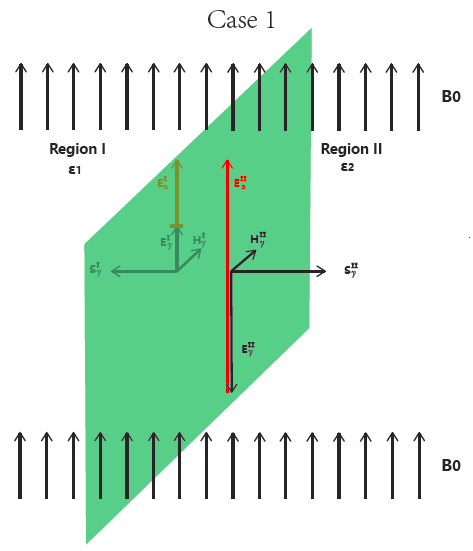}
\includegraphics[scale=0.6]{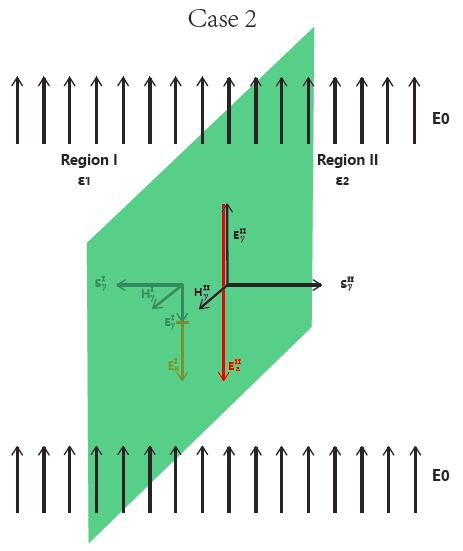}\\
\includegraphics[scale=0.6]{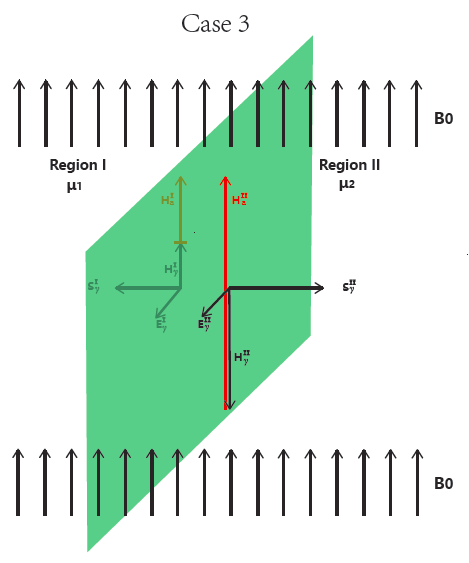}
\includegraphics[scale=0.6]{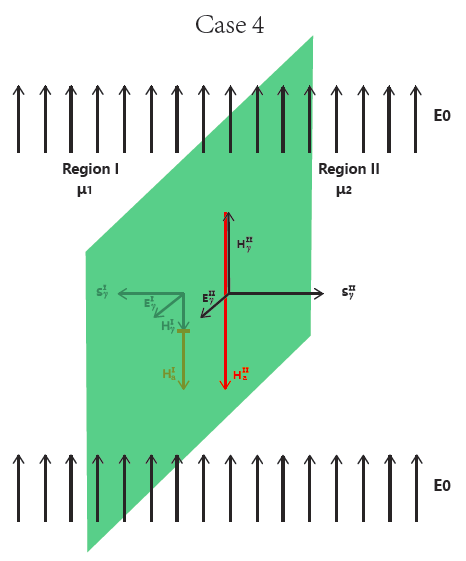}
\end{center}
\caption{The induced electromagnetic fields in two regions for four cases.}
\label{fig4case}
\end{figure}

We set up a configuration of interface between two regions $I$ and $II$ with a parallel static electromagnetic field $\vec{B}_0$ or $\vec{E}_0$.
The two regions are filled by media with different dielectric constant $\epsilon$ or magnetic permeability $\mu$.
As the propagating waves and the axion-induced electromagnetic fields are all parallel to the interface plane, we have the following continuity requirements between the two different regions
\begin{eqnarray}
\vec{E}_{\parallel}^I=\vec{E}_{\parallel}^{II}\;,~~\vec{H}_{\parallel}^I=\vec{H}_{\parallel}^{II}\;.
\end{eqnarray}
The general form of continuity in terms of the electromagnetic fields is
\begin{eqnarray}
\vec{E}_\gamma^I+\vec{E}_a^I+\vec{E}_0^I = \vec{E}_\gamma^{II}+\vec{E}_a^{II}+\vec{E}_0^{II}\;,~~\vec{H}_\gamma^I+\vec{H}_a^I+\vec{H}_0^I = \vec{H}_\gamma^{II}+\vec{H}_a^{II}+\vec{H}_0^{II}\;,
\end{eqnarray}
where we include the external static electromagnetic fields to account for the possibility of their direction the same as others or sudden change at the interface. In Fig.~\ref{fig4case}, we show the induced electromagnetic fields in two regions for four cases.

In this section, we discuss the above equations in different setups of the media for the two regions between the interface and the axion-induced radiation in QEMD.

%%%%%%%%%%%%%%%%%%%%%
\subsection{The radiation at an interface between two regions}
\label{sec:radiation}
%%%%%%%%%%%%%%%%%%%%%

%%%%%%%%%%%%%%%%%%%%%%%%%%
\subsubsection{Case 1: $\vec{B}_0\neq 0$, $\vec{E}_0=0$, $\epsilon_1\neq\epsilon_2$ and $\mu=1$}
%%%%%%%%%%%%%%%%%%%%%%%%%%

In this case, there is no static charge density causing an external electric field and the material is purely dielectric in the two regions. After applying the ordinary equations for static fields, the macroscopic Maxwell equations in a Fourier expansion become
\begin{eqnarray}
&&\vec{k}\times \hat{\vec{H}}_{\gamma a}+\omega \hat{\vec{D}} + i\hat{\vec{j}}'_{e,f}=g_{aEE}\omega \hat{a} \vec{B}_0+g_{aEM}\vec{B}_0\times \vec{k} \hat{a}\;,\nonumber\\
&&\vec{k}\times \hat{\vec{E}}_{\gamma a}-\omega \hat{\vec{B}}_{\gamma a} =-g_{aMM}\vec{B}_0\times \vec{k} \hat{a}-g_{aEM}\omega \hat{a} \vec{B}_0\;,\nonumber\\
&&\vec{k}\cdot \hat{\vec{B}}_{\gamma a} = g_{aEM}\vec{k}\cdot \vec{B}_0 \hat{a}\;,\nonumber\\
&&\vec{k}\cdot (\omega\hat{\vec{D}} +i\hat{\vec{j}}'_{e,f}) = g_{aEE}\omega\vec{k}\cdot \vec{B}_0 \hat{a} \;.
\end{eqnarray}
As $\vec{E}_0=0$, we can define the following relation in the media
\begin{eqnarray}
\hat{\vec{H}}_{\gamma a}={\hat{\vec{B}}_{\gamma a}\over \mu}\;,~~~
\omega\hat{\vec{D}} +i\hat{\vec{j}}'_{e,f}=\omega\hat{\vec{D}}_{\gamma a} +i\hat{\vec{j}}'_{e,f}=\omega\epsilon \vec{E}_{\gamma a}\;,
\end{eqnarray}
where $\mu$ is the magnetic permeability and $\epsilon$ is the total effective dielectric permittivity taking into account all electric effects. Using these relations, we get the linearized macroscopic axion equations
\begin{eqnarray}
&&\vec{k}\times {\hat{\vec{B}}_{\gamma a}\over \mu}+\omega \epsilon \hat{\vec{E}}_{\gamma a} =g_{aEE}\omega \hat{a} \vec{B}_0+g_{aEM}\vec{B}_0\times \vec{k} \hat{a}\;,\nonumber
\label{eq:lin1}\\
&&\vec{k}\times \hat{\vec{E}}_{\gamma a}-\omega \hat{\vec{B}}_{\gamma a} =-g_{aMM}\vec{B}_0\times \vec{k} \hat{a}-g_{aEM}\omega \hat{a} \vec{B}_0\;,\nonumber
\label{eq:lin2}\\
&&\vec{k}\cdot \hat{\vec{B}}_{\gamma a} = g_{aEM}\vec{k}\cdot \vec{B}_0 \hat{a}\;,\nonumber
\label{eq:lin3}\\
&&\epsilon \vec{k}\cdot \hat{\vec{E}}_{\gamma a} = g_{aEE}\vec{k}\cdot \vec{B}_0 \hat{a} \;,
\label{eq:lin4}
\end{eqnarray}
where $\vec{k}$ is a general symbol of wave vector which should be specified for particular fields below.

Then, in the limit of $\vec{k}_a = 0$, we obtain the axion-induced electromagnetic fields
\begin{eqnarray}
\vec{E}_a (t) = {1\over \epsilon }g_{aEE}  \vec{B}_0 a(t)\;,~~\vec{B}_a (t) = \mu \vec{H}_a (t) = g_{aEM} \vec{B}_0 a(t)\;,
\end{eqnarray}
as well as the propagating wave equations
\begin{eqnarray}
&&\vec{k}_\gamma\times \vec{H}_\gamma + \omega\epsilon \vec{E}_\gamma = 0\;,\\
&&\vec{k}_\gamma\times \vec{E}_\gamma - \omega \vec{B}_\gamma = \vec{k}_\gamma\times \vec{E}_\gamma - \omega \mu \vec{H}_\gamma= 0\;,
\end{eqnarray}
where $\vec{k}_\gamma$ denotes the wave vector of a propagating electromagnetic field with $k_\gamma^2=n^2 \omega^2$ and $n^2=\epsilon \mu$.
Here and below, we work in the approximation of axion with zero velocity $\vec{k}_a=0$ and frequency $\omega_a\approx m_a$. The axion DM field can be given by
\begin{eqnarray}
a(t)=a_0 e^{-im_a t}\;,
\end{eqnarray}
where $a_0=\sqrt{2\rho_{\rm DM}}/m_a$ with $\rho_{\rm DM}=0.4~{\rm GeV}~{\rm cm}^{-3}$ being the local DM density.

For the case of different dielectrics with $\epsilon_1\neq \epsilon_2$, we obtain the axion-induced electromagnetic fields
\begin{eqnarray}
\vec{H}_a^I=\vec{H}_a^{II}={1\over \mu} g_{aEM} \vec{B}_0 a_0 e^{-im_at}\;,~~\vec{E}_a^{I(II)}={1\over \epsilon_{1(2)}} g_{aEE}  \vec{B}_0 a_0 e^{-im_at}\;,
\end{eqnarray}
and the field values from continuity conditions become
\begin{eqnarray}
&&E_\gamma^I+E_a^I+E_0^I = E_\gamma^{II}+E_a^{II}+E_0^{II}\;,\nonumber\\
%&&H_\gamma^I+H_0^I=H_\gamma^{II}+H_0^{II} \Rightarrow  -{\epsilon_1\over n_1}E_\gamma^I + {B_0^I\over \mu}={\epsilon_2\over n_2}E_\gamma^{II} + {B_0^{II}\over \mu}
&&H_\gamma^I=H_\gamma^{II} \Rightarrow  -{\epsilon_1\over n_1}E_\gamma^I ={\epsilon_2\over n_2}E_\gamma^{II} \;,
\end{eqnarray}
where $H_\gamma$ is produced in the direction perpendicular to $E_\gamma$.
Inserting the obtained axion-induced fields to the above two equations, the solutions of propagating waves in the two regions are
\begin{eqnarray}
&&E_\gamma^I=+\Big[E_a^{II}-E_a^I+E_0^{II}-E_0^I\Big]{\epsilon_2 n_1\over \epsilon_1 n_2+\epsilon_2 n_1}\;,\nonumber\\
&&E_\gamma^{II}=-\Big[E_a^{II}-E_a^I+E_0^{II}-E_0^I\Big]{\epsilon_1 n_2\over \epsilon_1 n_2+\epsilon_2 n_1}\;,\nonumber\\
&&H_\gamma^I=-\Big[E_a^{II}-E_a^I+E_0^{II}-E_0^I\Big]{\epsilon_1 \epsilon_2\over \epsilon_1 n_2+\epsilon_2 n_1}\;,\nonumber\\
&&H_\gamma^{II}=-\Big[E_a^{II}-E_a^I+E_0^{II}-E_0^I\Big]{\epsilon_1 \epsilon_2\over \epsilon_1 n_2+\epsilon_2 n_1}\;,
\end{eqnarray}
where
\begin{eqnarray}
\Big[\cdots \Big] =
\Big({1\over \epsilon_2}-{1\over \epsilon_1}\Big)g_{aEE} B_0 a_0\;.
\end{eqnarray}
Thus, in this case, the external static magnetic field $B_0$ can be set to measure $g_{aEE}$ coupling.

%%%%%%%%%%%%%%%%%%%%%%%
\subsubsection{Case 2: $\vec{E}_0\neq 0$, $\vec{B}_0=0$, $\epsilon_1\neq \epsilon_2$ and $\mu=1$}
%%%%%%%%%%%%%%%%%%%%%%%

In this case, after applying the ordinary equations for static fields, the macroscopic Maxwell equations in a Fourier expansion become
\begin{eqnarray}
&&\vec{k}\times \hat{\vec{H}}_{\gamma a}+\omega \hat{\vec{D}} + i\hat{\vec{j}}'_{e,f}=g_{aEE}\vec{E}_0\times \vec{k}\hat{a}-g_{aEM}\omega\vec{E}_0\hat{a}\;,\nonumber\\
&&\vec{k}\times \hat{\vec{E}}_{\gamma a}-\omega \hat{\vec{B}}_{\gamma a} =g_{aMM}\omega\vec{E}_0\hat{a}-g_{aEM} \vec{E}_0\times\vec{k}\hat{a}\;,\nonumber\\
&&\vec{k}\cdot \hat{\vec{B}}_{\gamma a} = -g_{aMM}\vec{k}\cdot \vec{E}_0 \hat{a}\;,\nonumber\\
&&\vec{k}\cdot (\omega\hat{\vec{D}} +i\hat{\vec{j}}'_{e,f}) = -g_{aEM}\omega\vec{k}\cdot \vec{E}_0 \hat{a} \;.
\end{eqnarray}
As $\dot{\vec{E}}_0=0$, we can also define the following relation in the media
\begin{eqnarray}
\hat{\vec{H}}_{\gamma a}={\hat{\vec{B}}_{\gamma a}\over \mu}\;,~~~
\omega\hat{\vec{D}} +i\hat{\vec{j}}'_{e,f}=\omega\hat{\vec{D}}_{\gamma a} +i\hat{\vec{j}}'_{e,f}=\omega\epsilon \vec{E}_{\gamma a}\;.
\end{eqnarray}
Then, we get the linearized macroscopic axion equations
\begin{eqnarray}
&&\vec{k}\times {\hat{\vec{B}}_{\gamma a}\over \mu}+\omega \epsilon \hat{\vec{E}}_{\gamma a} =g_{aEE}\vec{E}_0\times \vec{k}\hat{a}-g_{aEM}\omega\vec{E}_0\hat{a}\;,\nonumber
\label{eq:lin1}\\
&&\vec{k}\times \hat{\vec{E}}_{\gamma a}-\omega \hat{\vec{B}}_{\gamma a} =g_{aMM}\omega\vec{E}_0\hat{a}-g_{aEM} \vec{E}_0\times\vec{k}\hat{a}\;,\nonumber
\label{eq:lin2}\\
&&\vec{k}\cdot \hat{\vec{B}}_{\gamma a} = -g_{aMM}\vec{k}\cdot \vec{E}_0 \hat{a}\;,\nonumber
\label{eq:lin3}\\
&&\epsilon \vec{k}\cdot \hat{\vec{E}}_{\gamma a} = -g_{aEM}\vec{k}\cdot \vec{E}_0 \hat{a} \;.
\label{eq:lin4}
\end{eqnarray}
In the limit of $\vec{k}_a = 0$, we obtain the axion-induced electromagnetic fields
\begin{eqnarray}
\vec{E}_a (t) =- {1\over \epsilon }g_{aEM}  \vec{E}_0 a(t)\;,~~\vec{B}_a (t) = \mu \vec{H}_a (t) = -g_{aMM} \vec{E}_0 a(t)\;,
\end{eqnarray}
as well as the propagating wave equations also apply here.

For the case of different dielectrics with $\epsilon_1\neq \epsilon_2$, we obtain the axion-induced electromagnetic fields
\begin{eqnarray}
\vec{H}_a^I=\vec{H}_a^{II}=-{1\over \mu} g_{aMM} \vec{E}_0 a_0 e^{-im_at}\;,~~\vec{E}_a^{I(II)}=-{1\over \epsilon_{1(2)}} g_{aEM}  \vec{E}_0 a_0 e^{-im_at}\;,
\end{eqnarray}
and the field values from continuity conditions become
\begin{eqnarray}
&&E_\gamma^I+E_a^I+E_0^I = E_\gamma^{II}+E_a^{II}+E_0^{II}\;,\nonumber\\
%&&H_\gamma^I+H_0^I=H_\gamma^{II}+H_0^{II} \Rightarrow  -{\epsilon_1\over n_1}E_\gamma^I + {B_0^I\over \mu}={\epsilon_2\over n_2}E_\gamma^{II} + {B_0^{II}\over \mu}\;,\\
&&H_\gamma^I=H_\gamma^{II} \Rightarrow  -{\epsilon_1\over n_1}E_\gamma^I ={\epsilon_2\over n_2}E_\gamma^{II} \;.
\end{eqnarray}
Inserting the obtained axion-induced fields to the above two equations, the solutions of propagating waves in the two regions are
\begin{eqnarray}
&&E_\gamma^I=+\Big[E_a^{II}-E_a^I+E_0^{II}-E_0^I\Big]{\epsilon_2 n_1\over \epsilon_1 n_2+\epsilon_2 n_1}\;,\nonumber\\
&&E_\gamma^{II}=-\Big[E_a^{II}-E_a^I+E_0^{II}-E_0^I\Big]{\epsilon_1 n_2\over \epsilon_1 n_2+\epsilon_2 n_1}\;,\nonumber\\
&&H_\gamma^I=-\Big[E_a^{II}-E_a^I+E_0^{II}-E_0^I\Big]{\epsilon_1 \epsilon_2\over \epsilon_1 n_2+\epsilon_2 n_1}\;,\nonumber\\
&&H_\gamma^{II}=-\Big[E_a^{II}-E_a^I+E_0^{II}-E_0^I\Big]{\epsilon_1 \epsilon_2\over \epsilon_1 n_2+\epsilon_2 n_1}\;,
\end{eqnarray}
where
\begin{eqnarray}
\Big[\cdots \Big] = -
\Big({1\over \epsilon_2}-{1\over \epsilon_1}\Big)g_{aEM} E_0 a_0\;.
\end{eqnarray}
Thus, in this case, the external static electric field $E_0$ can be set to measure $g_{aEM}$ coupling.

%%%%%%%%%%%%%%%%%%%%%%%%%%%%
\subsubsection{Case 3: $\vec{B}_0\neq 0$, $\vec{E}_0=0$, $\mu_1\neq\mu_2$ and $\epsilon=1$}
%%%%%%%%%%%%%%%%%%%%%%%%%%%%

In this case, we obtain
the macroscopic Maxwell equations in a Fourier expansion as
\begin{eqnarray}
&&\vec{k}\times \hat{\vec{H}}_{\gamma a}+\omega \hat{\vec{D}}_{\gamma a} + i\hat{\vec{j}}'_{e,f}=g_{aEE}\omega \hat{a} \vec{B}_0+g_{aEM}\vec{B}_0\times \vec{k} \hat{a}\;,\nonumber\\
&&\vec{k}\times \hat{\vec{E}}_{\gamma a}-\omega \hat{\vec{B}}_{\gamma a} =-g_{aMM}\vec{B}_0\times \vec{k} \hat{a}-g_{aEM}\omega \hat{a} \vec{B}_0\;,\nonumber\\
&&\vec{k}\cdot \hat{\vec{B}}_{\gamma a} = g_{aEM}\vec{k}\cdot \vec{B}_0 \hat{a}\;,\nonumber\\
&&\vec{k}\cdot (\omega\hat{\vec{D}}_{\gamma a}+i\hat{\vec{j}}'_{e,f}) = g_{aEE}\omega\vec{k}\cdot \vec{B}_0 \hat{a} \;.
\end{eqnarray}
They are exactly the same as those in case 1 and the solutions of $\vec{E}_a$ and $\vec{B}_a$ are also the same.

For the case of different magnetic materials with $\mu_1\neq \mu_2$ and $\epsilon_1=\epsilon_2=\epsilon$, we have the axion-induced electromagnetic fields
\begin{eqnarray}
\vec{E}_a^{I}=\vec{E}_a^{II}={1\over \epsilon} g_{aEE}  \vec{B}_0 a_0 e^{-im_at}\;,~~\vec{H}_a^{I(II)}={1\over \mu_{1(2)}} g_{aEM} \vec{B}_0 a_0 e^{-im_at}\;,
\end{eqnarray}
and the field values from continuity conditions are
\begin{eqnarray}
&&H_\gamma^I+H_a^I+H_0^I = H_\gamma^{II}+H_a^{II}+H_0^{II}\;,\nonumber\\
&&E_\gamma^I=E_\gamma^{II} \Rightarrow  -{\mu_1\over n_1}H_\gamma^I ={\mu_2\over n_2}H_\gamma^{II} \;.
\end{eqnarray}
The solutions of propagating waves in the two regions are
\begin{eqnarray}
&&H_\gamma^I=+\Big[H_a^{II}-H_a^I+{B_0^{II}\over \mu_2}-{B_0^I\over \mu_1}\Big]{\mu_2 n_1\over \mu_1 n_2+\mu_2 n_1}\;,\nonumber\\
&&H_\gamma^{II}=-\Big[H_a^{II}-H_a^I+{B_0^{II}\over \mu_2}-{B_0^I\over \mu_1}\Big]{\mu_1 n_2\over \mu_1 n_2+\mu_2 n_1}\;,\nonumber\\
&&E_\gamma^I=-\Big[H_a^{II}-H_a^I+{B_0^{II}\over \mu_2}-{B_0^I\over \mu_1}\Big]{\mu_1 \mu_2\over \mu_1 n_2+\mu_2 n_1}\;,\nonumber\\
&&E_\gamma^{II}=-\Big[H_a^{II}-H_a^I+{B_0^{II}\over \mu_2}-{B_0^I\over \mu_1}\Big]{\mu_1 \mu_2\over \mu_1 n_2+\mu_2 n_1}\;,
\end{eqnarray}
where
\begin{eqnarray}
\Big[\cdots \Big] =
    \Big({1\over \mu_2}-{1\over \mu_1}\Big) \Big(g_{aEM} B_0 a_0+B_0 \Big)\approx \Big({1\over \mu_2}-{1\over \mu_1}\Big)B_0\;.
\end{eqnarray}
In this case, the propagating waves are approximately independent of the axion coupling and thus cannot be used to measure it because the external magnetic field is embedded in the $H$ continuity condition.

%%%%%%%%%%%%%%%%%%%%%%%%%%%%%%%%%
\subsubsection{Case 4: $\vec{E}_0\neq 0$, $\vec{B}_0=0$, $\mu_1\neq \mu_2$ and $\epsilon=1$}
%%%%%%%%%%%%%%%%%%%%%%%%%%%%%%%%%%

In this case,
the macroscopic Maxwell equations in a Fourier expansion are
\begin{eqnarray}
&&\vec{k}\times \hat{\vec{H}}_{\gamma a}+\omega \hat{\vec{D}}_{\gamma a} + i\hat{\vec{j}}'_{e,f}=g_{aEE}\vec{E}_0\times \vec{k}\hat{a}-g_{aEM}\omega\vec{E}_0\hat{a}\;,\nonumber\\
&&\vec{k}\times \hat{\vec{E}}_{\gamma a}-\omega \hat{\vec{B}}_{\gamma a} =g_{aMM}\omega\vec{E}_0\hat{a}-g_{aEM} \vec{E}_0\times\vec{k}\hat{a}\;,\nonumber\\
&&\vec{k}\cdot \hat{\vec{B}}_{\gamma a} = -g_{aMM}\vec{k}\cdot \vec{E}_0 \hat{a}\;,\nonumber\\
&&\vec{k}\cdot (\omega\hat{\vec{D}}_{\gamma a} +i\hat{\vec{j}}'_{e,f}) = -g_{aEM}\omega\vec{k}\cdot \vec{E}_0 \hat{a} \;.
\end{eqnarray}
They are exactly the same as those in case 2 and the solutions of $\vec{E}_a$ and $\vec{B}_a$ are also the same.

The axion-induced electromagnetic fields are then
\begin{eqnarray}
\vec{E}_a^{I}=\vec{E}_a^{II}=-{1\over \epsilon} g_{aEM}  \vec{E}_0 a_0 e^{-im_at}\;,~~\vec{H}_a^{I(II)}=-{1\over \mu_{1(2)}} g_{aMM} \vec{E}_0 a_0 e^{-im_at}\;.
\end{eqnarray}
and the field values from continuity conditions are
\begin{eqnarray}
&&H_\gamma^I+H_a^I+H_0^I = H_\gamma^{II}+H_a^{II}+H_0^{II}\;,\nonumber\\
&&E_\gamma^I=E_\gamma^{II}\Rightarrow  -{\mu_1\over n_1}H_\gamma^I ={\mu_2\over n_2}H_\gamma^{II} \;.
\end{eqnarray}
The solutions of propagating waves in the two regions are
\begin{eqnarray}
&&H_\gamma^I=+\Big[H_a^{II}-H_a^I+{B_0^{II}\over \mu_2}-{B_0^I\over \mu_1}\Big]{\mu_2 n_1\over \mu_1 n_2+\mu_2 n_1}\;,\nonumber\\
&&H_\gamma^{II}=-\Big[H_a^{II}-H_a^I+{B_0^{II}\over \mu_2}-{B_0^I\over \mu_1}\Big]{\mu_1 n_2\over \mu_1 n_2+\mu_2 n_1}\;,\nonumber\\
&&E_\gamma^I=-\Big[H_a^{II}-H_a^I+{B_0^{II}\over \mu_2}-{B_0^I\over \mu_1}\Big]{\mu_1 \mu_2\over \mu_1 n_2+\mu_2 n_1}\;,\nonumber\\
&&E_\gamma^{II}=-\Big[H_a^{II}-H_a^I+{B_0^{II}\over \mu_2}-{B_0^I\over \mu_1}\Big]{\mu_1 \mu_2\over \mu_1 n_2+\mu_2 n_1}\;,
\end{eqnarray}
where
\begin{eqnarray}
\Big[\cdots \Big] =
    -\Big({1\over \mu_2}-{1\over \mu_1}\Big) g_{aMM} E_0 a_0\;.
\end{eqnarray}
Thus, in this case, the external static electric field $E_0$ can be set to measure $g_{aMM}$ coupling.

%%%%%%%%%%%%%%%%%%%%%
\subsection{Poynting's theorem and the energy flux density}
\label{sec:energyflux}
%%%%%%%%%%%%%%%%%%%%%

The energy flux density of an EM propagating field  is given by the Poynting’s theorem as
\begin{align}
\vec{S}_\gamma=\vec{E}_\gamma\times\vec{H}_\gamma\;.
\end{align}
Next, we give the results of energy flux density for the above viable cases (case 1, 2 and 4).

The propagating waves in Case 1 are
\begin{align}
E_\gamma^I&=g_{aEE} B_0 a_0 \frac{\epsilon _1-\epsilon_2}{\epsilon _1 \sqrt{\epsilon _2}+\epsilon_2 \sqrt{\epsilon _1}}\frac{1}{\sqrt{\epsilon _1}}\;,
\nonumber\\
E_\gamma^{II}&=-g_{aEE} B_0 a_0 \frac{\epsilon_1-\epsilon_2}{\epsilon_1 \sqrt{\epsilon _2}+\epsilon_2 \sqrt{\epsilon _1}}\frac{1}{\sqrt{\epsilon_ 2}}\;,
\nonumber\\
H_\gamma^I=H_\gamma^{II}&=-g_{aEE} B_0 a_0 \frac{\epsilon_1-\epsilon_2}{\epsilon_1 \sqrt{\epsilon_2}+\epsilon_2 \sqrt{\epsilon_1}}\;,
\end{align}
where $\epsilon_1=n_1^2$ and $\epsilon_2=n_2^2$.
After plugging them into the Poynting's theorem, we obtain
\begin{align}
\overline{S}^i_{\gamma}=\mp \frac{1}{\sqrt{\epsilon _i}}\frac{(g_{aEE} B_0 a_0)^2}{2}\Big(\frac{\epsilon_1-\epsilon_2}{\epsilon_1\sqrt{\epsilon_2}+\epsilon_2\sqrt{\epsilon_1}}\Big)^2=\mp \frac{1}{\sqrt{\epsilon _i}}\frac{(g_{aEE} B_0 a_0)^2}{2}\Big( {1\over \sqrt{\epsilon_2}}-{1\over \sqrt{\epsilon_1}} \Big)^2\;,
\end{align}
where $i=I$ (upper sign) or $II$ (lower sign), and we take into account a factor of $1/2$ to obtain the time-averaged value.
For Case 2, one just needs to make a replacement of $g_{aEE}B_0\to -g_{aEM}E_0$ in Case 1's results.

The propagating waves in Case 4 are
\begin{align}
H_\gamma^I&=-g_{aMM} E_0 a_0 \frac{\mu _1-\mu_2}{\mu _1 \sqrt{\mu _2}+\mu _2 \sqrt{\mu _1}}\frac{1}{\sqrt{\mu _1}}\;,
\nonumber\\
H_\gamma^{II}&=g_{aMM} E_0 a_0 \frac{\mu_1-\mu_2}{\mu_1 \sqrt{\mu _2}+\mu_2 \sqrt{\mu _1}}\frac{1}{\sqrt{\mu_ 2}}\;,
\nonumber\\
E_\gamma^I=E_\gamma^{II}&=g_{aMM} E_0 a_0 \frac{\mu_1-\mu_2}{\mu_1 \sqrt{\mu_2}+\mu_2\sqrt{\mu_1}}\;.
\end{align}
After plugging them into the Poynting's theorem, we obtain
\begin{align}
\overline{S}^i_{\gamma}=\mp \frac{1}{\sqrt{\mu _i}}\frac{(g_{aMM} E_0 a_0)^2}{2}\Big(\frac{\mu_1-\mu_2}{\mu_1\sqrt{\mu_2}+\mu_2\sqrt{\mu_1}}\Big)^2=\mp \frac{1}{\sqrt{\mu_i}}\frac{(g_{aMM} E_0 a_0)^2}{2}\Big( {1\over \sqrt{\mu_2}}-{1\over \sqrt{\mu_1}} \Big)^2\;.
\end{align}

%%%%%%%%%%%%%%%%%%%%%%%%%%%%%%%%%%%%%%%%%
\section{Sensitivity of interface haloscopes to new axion couplings in QEMD}
\label{sec:Exp}
%%%%%%%%%%%%%%%%%%%%%%%%%%%%%%%%%%%%%%%%%

In this section, based on the above energy flux density and signal-to-noise ratio (SNR), we calculate the sensitivity of interface haloscopes to the new axion couplings in QEMD. Following Ref.~\cite{Millar:2016cjp}, we assume an extreme case with medium II being vacuum ($n_2=1$) and thus take the interface as a perfect mirror to emit radiation into vacuum region II. Moreover, as pointed out by Ref.~\cite{Millar:2016cjp}, the axion-induced electromagnetic wave can be boosted with a series of parallel interfaces between different media.
The outgoing wave then becomes a coherent superposition of the transmission and reflection at each interface. We follow Ref.~\cite{Caldwell:2016dcw,Millar:2016cjp} to choose an experimental setup of 80 disks and take a boost factor $\beta$ to describe this effect. The signal power can be given by
\begin{align}
P_{\rm signal}= A \beta^2 \eta \overline{S}^{II}_{\gamma}\;,
\end{align}
where $\eta=0.8$ is power efficiency, and $A=1~{\rm m}^2$ is disk area.
To measure this signal, we choose HEMT amplifiers with $T_{\rm sys}=8~{\rm K}$ and the estimated  noise is
\begin{align}
P_{\rm noise}=T_{\rm sys}\sqrt{\frac{\Delta\nu}{\Delta t}}\;,
\end{align}
where a suitable measurement time is $\Delta t=1$ day, and a signal bandwidth is given by the frequency range $\Delta \nu=\Delta \omega_a/2\pi\sim 10^{-6} m_a/2\pi$.
$\beta$ is obtained from Area Law: $\beta^2 \Delta \nu_{\beta}=K$ with $K$ being a constant associated with experimental setup~\cite{Millar:2016cjp}.

\begin{figure}[htb!]
\centering
\includegraphics[scale=0.22]{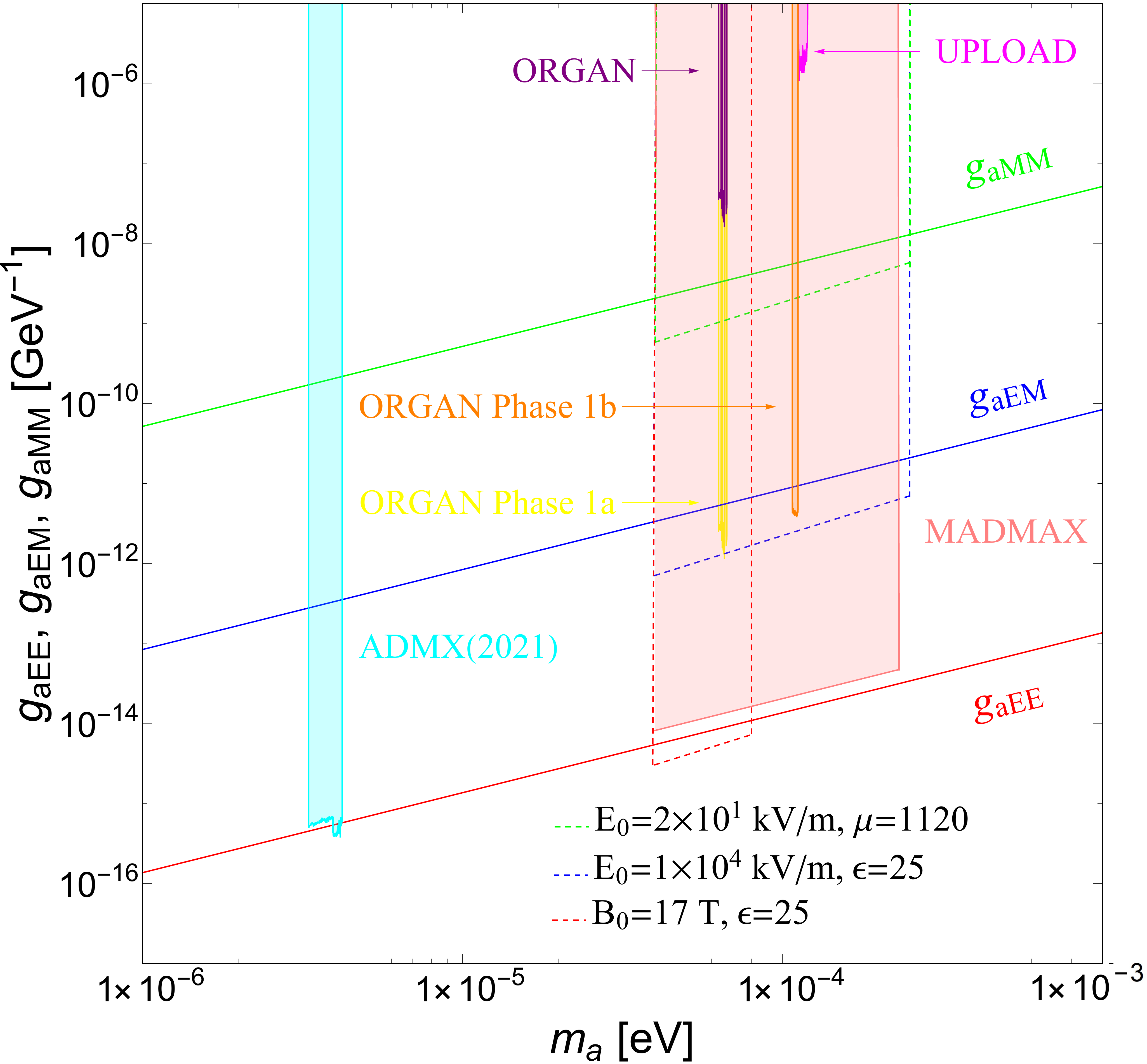}
\caption{The expected sensitivity bounds of $g_{aEE}$ (red dashed line), $g_{aEM}$ (blue dashed line) and $g_{aMM}$ (green dashed line). $\beta\simeq 275$ for $g_{aEM},g_{aMM}$ and $\beta\simeq 628$ for $g_{aEE}$. The theoretical predictions of $g_{aEE}$, $g_{aEM}$ and $g_{aMM}$ (solid) are also presented. Some existing or potential exclusion limits on conventional coupling $g_{a\gamma\gamma}$ are shown for reference, including ADMX (cyan)~\cite{ADMX:2021nhd}, ORGAN Phase 1a (yellow)~\cite{Quiskamp:2022pks} and Phase 1b (orange)~\cite{Quiskamp:2023ehr}, and the proposed MADMAX (pink)~\cite{Caldwell:2016dcw}. The existing limits on $g_{aMM}$ from upconversion experiment UPLOAD (magenta)~\cite{Thomson:2023moc} and $g_{aEM}$ from ORGAN experiment (purple)~\cite{McAllister:2022ibe} are also shown for comparison.
}
\label{fig:sensitivity}
\end{figure}

The SNR is required to satisfy
\begin{align}
\frac{P_{\rm signal}}{P_{\rm noise} }>5\;.
\end{align}
The total run time is given by $t_{\rm tot}=(\Delta t +t_R)\frac{\Delta m_a}{\Delta \nu_{\beta}}$ for a suitable range of axion mass $\Delta m_a$ starting  from $40~\mu {\rm eV}$. We take $t_R=1$ day as a conservative estimated value of readjust time before each detection and $\Delta \nu_{\beta}=50~{\rm MHz}$ with $t_{\rm tot}=5$ years as a reasonable choice to balance the total run time and detection range. This choice results in $\beta\simeq \mathcal{O}(100)$~\cite{Caldwell:2016dcw,Millar:2016cjp}. We keep this $\beta$ value unchanged in whole experimental process, and then plug the above parameters into SNR formula to obtained the sensitivity bound of new axion couplings. In Fig.~\ref{fig:sensitivity}, after taking a few experimental setups for illustration, we show the sensitivity bound of couplings $g_{aEE}$, $g_{aEM}$ and $g_{aMM}$ together with their theoretical predictions. We find that the reasonable setups of interface haloscopes with perfect mirror can probe the theoretical predictions of $g_{aEE}$, $g_{aEM}$ and $g_{aMM}$ for $\mathcal{O}(10)~\mu{\rm eV}\lesssim m_a \lesssim \mathcal{O}(100)~\mu{\rm eV}$.
It turns out that a large background electric field is needed to probe $g_{aEM}$ and $g_{aMM}$. The large static electric field could induce
extra currents/charges in the detector volume depending on the design and therefore may influence the sensitivity. We leave the experimental design to future study.

%%%%%%%%%%%%%%%%%%%%%%%%%%%%%%%%
\section{Conclusion}
\label{sec:Con}
%%%%%%%%%%%%%%%%%%%%%%%%%%%%%%%%%

The so-called
Witten effect implies the relationship between axion and magnetic monopole. A sound QFT theory called QEMD in the presence of magnetic monopoles was utilized to construct a more generic axion-photon Lagrangian in the low-energy axion effective field theory. This generic axion-photon Lagrangian introduces the interactions between axion and two four-potentials. It leads to new axion-modified Maxwell equations. To search for high-mass axions $m_a\gtrsim \mathcal{O}(10)~\mu{\rm eV}$, the interface haloscopes were proposed by placing an interface between two electromagnetic media with different properties.

In this work, for the generic axion-photon couplings built under QEMD, we provide comprehensive expressions for the axion-induced electromagnetic fields and propagating waves in different setups of interface and background fields. We also calculate energy flux densities, the signal-to-noise ratio and obtain the sensitivity to new axion-photon couplings ($g_{aEE}$, $g_{aEM}$ and $g_{aMM}$) for high-mass axions. We find
\begin{itemize}
\item The configuration of interface between two dielectric regions and a parallel static magnetic (electric) field can measure $g_{aEE}$ ($g_{aEM}$) coupling.
\item The configuration of interface between two regions with magnetic material and a parallel static electric field can measure $g_{aMM}$ coupling.
\item A reasonable setup of interface haloscopes with perfect mirror can probe the theoretical predictions of $g_{aEE}$, $g_{aEM}$ and $g_{aMM}$ for $\mathcal{O}(10)~\mu{\rm eV}\lesssim m_a \lesssim \mathcal{O}(100)~\mu{\rm eV}$.
\end{itemize}

%###################################################################
%%%%%%%%%%%%%%%%%%%%%%%%
\acknowledgments
%%%%%%%%%%%%%%%%%%%%%%%%%%%%%%%%%%%%%%%%%%%%
We thank Anton V.~Sokolov and Andreas Ringwald for useful comment. T.L. is supported by the National Natural Science Foundation of China (Grant No. 12375096, 12035008, 11975129) and ``the Fundamental Research Funds for the Central Universities'', Nankai University (Grants No. 63196013).

%%%%%%%%%%%%%%%%%%%%%%%%%%%%%%%%%%%%%%%%%%%%%%
%%%%%%%%%%%%%%%%%%%%%%%%%%%%%%%%%%%%%%%%%%%%%%
%%%%%%%%%%%%%%%%%%%%%%%%%%%%%%%%%%%%%
%\bibliographystyle{JHEP}
\bibliography{refs}

\end{document}